\newcommand{\U}[1]{\,{\rm #1}}

\documentclass[10pt,twocolumns]{iopart}

\usepackage{graphicx}
\usepackage{bm}
\usepackage{multirow}
\usepackage{iopams}

\begin{document}

\title[]{Experimental observations and modelling of intrinsic rotation reversals in tokamaks}
\author{Y.~Camenen$^1$, C.~Angioni$^2$, A.~Bortolon$^3$, B.P.~Duval$^4$, E.~Fable$^2$, W.A.~Hornsby$^2$, R.M.~McDermott$^2$, D.H. Na$^5$, Y-S. Na$^5$, A.G.~Peeters$^6$ and J.E.~Rice$^7$}
\address{$^1$ CNRS, Aix-Marseille Univ., PIIM UMR7345, Marseille, France}
\address{$^2$ Max Planck Institut f\"{u}r Plasmaphysik, Garching, Germany}
\address{$^3$ Princeton Plasma Physics Laboratory, Princeton, USA}
\address{$^4$ EPFL, Swiss Plasma Center (SPC), Lausanne, Switzerland}
\address{$^5$ Departement of Nuclear Engineering, Seoul National University, Seoul, Korea}
\address{$^6$ Physics Department, University of Bayreuth, Bayreuth, Germany}
\address{$^7$ PSFC, MIT, Cambridge, Massachusetts, USA}

\begin{abstract}
The progress made in understanding spontaneous toroidal rotation reversals in tokamaks is reviewed and current ideas to solve this ten-year-old puzzle are explored. The paper includes a summarial synthesis of the experimental observations in AUG, C-Mod, KSTAR, MAST and TCV tokamaks, reasons why turbulent momentum transport is thought to be responsible for the reversals,  a review of the theory of turbulent momentum transport and suggestions for future investigations.
\end{abstract}

\ioptwocol

\section{Introduction}
During the 1980s, it was shown that stationary toroidal flows that reach up to 20$\%$ of the thermal velocity can develop in tokamak plasmas in the absence of externally applied torque  (a summary of these early observations is available in Table~1 of \cite{Rice:NF97}). 
This phenomenon, dubbed \textit{intrinsic rotation}, has practical implications for future low torque devices like ITER owing to the potential stabilising impact of plasma flows on turbulence and deleterious MHD instabilities. 

Following initial measurements, experiments were performed to explore the physics of intrinsic rotation. The observations have been regularly summarised in review articles \cite{Ida:PPCF98,Rice:JPCS08,DeGrassie:PPCF09,Ida:NF14,Rice:PPCF16}.
Momentum transport, reconnection events (sawteeth and ELMs), non-axisymmetric magnetic fields (from magnetic perturbation coils, error fields or large MHD modes), orbit losses and interactions with neutrals were all observed to impact intrinsic rotation. The underlying physics is surprisingly rich, involving many competing mechanisms that determine the final rotation profile. These include collisional momentum transport, fluctuation-induced Reynolds and Maxwell stresses (turbulent transport), charge exchange with neutrals, $J_\parallel\times \delta B$ torque induced by resonant magnetic perturbations,  $J_r\times B$ torque induced by non-axisymmetric magnetic fields (neoclassical toroidal viscosity NTV), ionisation currents and orbit losses. A convenient set of transport equations has been proposed in \cite{Callen:PoP09} to describe the evolution of toroidal flows in tokamak plasmas resulting from these various mechanisms in a consistent fluid moment framework. This approach highlights the complexity of intrinsic rotation inherent to the number of mechanisms at play.

The present paper focuses on a specific puzzle within intrinsic rotation: spontaneous toroidal rotation reversals. 
This intriguing phenomenon was reported 10 years ago on the TCV tokamak where the core toroidal rotation was observed to flip from counter-current to co-current when a threshold in density was exceeded in Ohmic L-mode plasmas \cite{Bortolon:PRL06}. Rotation reversals have since been demonstrated in C-Mod \cite{Rice:PPCF08}, AUG \cite{McDermott:NF14}, MAST \cite{Hillesheim:NF15} and KSTAR \cite{Na:NF16}. 
In parallel, the theory of intrinsic rotation has undergone considerable development and many possible physical mechanisms have been identified.
In spite of this progress, understanding toroidal rotation reversals still eludes us and predicting the direction of the core rotation in Ohmic L-modes remains a challenge. Toroidal rotation reversals do not directly affect plasma performance, but they represent a critical test for the theory of intrinsic rotation. The purpose of the present work is to survey the observations and the theoretical framework with the goal of presenting the current understanding of this research and explaining current ideas and approaches to its resolution.

The definitions and conventions adopted in this paper are introduced in Sec.~\ref{sec:def}, followed by a summary of the experimental observations in Sec.~\ref{sec:expobs}. The constraints these observations put on the theory and, in particular, the reasons why turbulent momentum transport is thought to be responsible for rotation reversals are discussed in Sec.~\ref{sec:inteprexp}.
The theory of turbulent momentum transport is briefly reviewed in Sec.~\ref{sec:theory} before summarising the current status of the modelling activities in Sec.~\ref{sec:modelling}. Finally, in Sec.~\ref{sec:discussion} future work and open issues are discussed.

\section{Definitions and conventions} \label{sec:def}
Throughout this paper \textit{intrinsic rotation} refers to the toroidal rotation that develops in the absence of externally applied torque. The toroidal rotation is noted $v_\varphi$ and its direction is given with respect to the plasma current: $v_\varphi>0$ for co-current rotation and $v_\varphi<0$ for counter-current rotation. 
The rotation profile is said to be peaked (hollow) when  $v_\varphi$ increases (decreases) from the edge to the magnetic axis. Note that other definitions exist in the litterature, where for instance $|v_\varphi|$ is used in place of $v_\varphi$ to define peaked and hollow profiles. The present definition is deemed more appropriate to describe profiles that cross $v_\varphi=0$.
Three regions are distinguished in the rotation profile following \cite{Duval:PPCF07, Duval:PoP08, Bortolon:PhD09}: the sawtooth region $0\leq r/a \lesssim r_{\rm inv}/a$, with $r_{\rm inv}$ the sawtooth inversion radius and $a$ the plasma minor radius, the gradient region $r_{\rm inv}/a \lesssim r/a \lesssim 0.8$ (typically) and the edge region $0.8 \lesssim r/a \leq 1$. The separation between the gradient and edge regions relies on the different dependencies of the toroidal rotation gradient on plasma parameters in these two regions.  The plasma core includes the sawtooth and gradient regions.

\textit{Toroidal rotation reversals} are defined as a large change ($\gtrsim 100\%$) of the intrinsic toroidal rotation gradient over the whole gradient region triggered by minor changes ($\lesssim 20\%$) in the control plasma parameters. 
It is important to notice that this definition does not require a change in the direction of the central toroidal rotation, nor of the toroidal rotation gradient, which is somehow at odds with the \textit{reversal} qualifier. \textit{Toroidal rotation bifurcation} would be a more accurate description of this phenomenology. For consistency with the past literature, however, we retain \textit{toroidal rotation reversal} and broaden its definition to include cases where the sign of the toroidal rotation and/or of its gradient do not change.

\section{Experimental observations} \label{sec:expobs}
The experimental observations of toroidal rotation reversals collected over the last ten years in AUG \cite{McDermott:NF14, Angioni:PRL11}, C-Mod \cite{Rice:NF11, Rice:PRL11b, Rice:PoP12, Rice:NF13, Reinke:PPCF13}, MAST \cite{Hillesheim:NF15}, KSTAR \cite{Na:NF16} and TCV \cite{Bortolon:PRL06, Duval:PPCF07, Duval:PoP08, Bortolon:PhD09, Sauter:IAEA10} are summarised in this section. In some cases, direct reference is made to figures in the published works.

\subsection{Measurements}
\begin{enumerate}
\item Toroidal rotation reversals were reported observing impurity ions (boron, carbon and argon) with a variety of diagnostics: X-ray imaging crystal spectrocsopy (XICS) in C-Mod and KSTAR (argon), charge exchange recombination spectroscopy (CXRS) with a diagnostic neutral beam in TCV (carbon), CXRS using short pulses of a heating neutral beam in AUG (boron) and KSTAR (carbon).
\item Reversals have also been inferred from Doppler back-scattering measurements of the perpendicular velocity of electron density fluctuations in AUG and MAST (assuming a dominant $E\times B$ velocity). 
\item The measurements were mostly performed in Ohmic L-modes, but reversals were also reported in the presence of ion cyclotron heating \cite{Reinke:PPCF13} and electron cyclotron heating \cite{Bortolon:PhD09}, still in L-mode. 
\end{enumerate}

\subsection{Triggers}
\begin{enumerate}
\item Toroidal rotation reversals have been triggered by density ramps, plasma current-ramps, toroidal magnetic field ramps, impurity injection and by switching on/off electron cyclotron heating, see e.g. \cite{Rice:PPCF16, Bortolon:PhD09}.
\item The reversals appear to be highly reproducible and weakly sensitive to machine conditioning \cite{Duval:PPCF07}. 
\end{enumerate}

\subsection{Reversal direction}
The reversal direction is discussed here as a function of increasing density, as density ramps are the most common trigger of reversals.
\begin{enumerate}
\item Type I. Co-current to counter-current reversals (or more precisely bifurcations from peaked/flat to hollow profiles in the gradient region) have been observed in AUG, C-Mod, MAST and TCV (diverted configuration) and in KSTAR (limited configuration) above a critical density.\\
{\footnotesize AUG: Fig. 4 in \cite{McDermott:NF14}, C-Mod Fig. 13 in \cite{Rice:NF11}, KSTAR: Fig.~15 in \cite{Na:NF16}, MAST: Fig. 4 in \cite{Hillesheim:NF15}, TCV: Fig. 6 in \cite{Duval:PPCF07}.}
\item Type II. Counter-current to co-current reversals (transition from hollow to peaked profiles) have been observed in TCV limited plasmas for $q_{95}\lesssim3$ (Type II.a reversals), in AUG and TCV diverted plasmas at very high density (Type II.b reversals) and in MAST low current and low density diverted plasmas (Type II.c reversals).\\ 
{\footnotesize AUG: Fig. 4 in \cite{Angioni:PRL11} and Fig. 7 in \cite{McDermott:NF14}, TCV: Fig. 1 in \cite{Bortolon:PRL06}}
\end{enumerate}
Note that the distinction made above is purely phenomenological and does not exclude that all reversals be manifestations of the same physical mechanism observed in different plasma conditions. 
In particular, in Ohmic plasmas, $T_e$, $T_i$, $n_e$ and $q$ are strongly coupled and a unique threshold identified as a combination of these parameters may be traversed several times in a density ramp, with a trajectory possibly dependent on the operation mode (limited versus diverted for instance).

\subsection{Initial and final states}
\begin{figure*}[!htb]
        \begin{center}
        \includegraphics[width=0.32\textwidth]{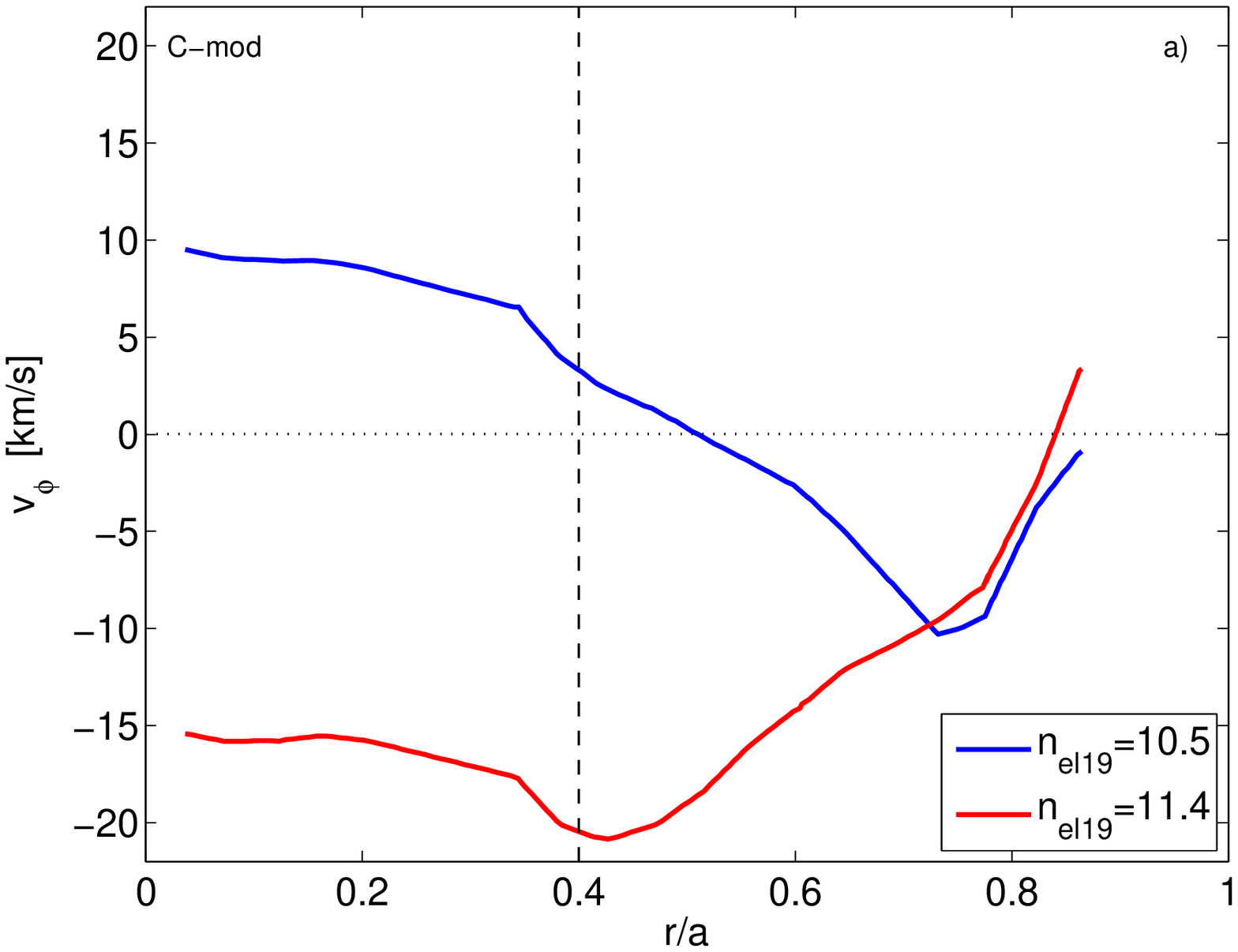}%
        \includegraphics[width=0.32\textwidth]{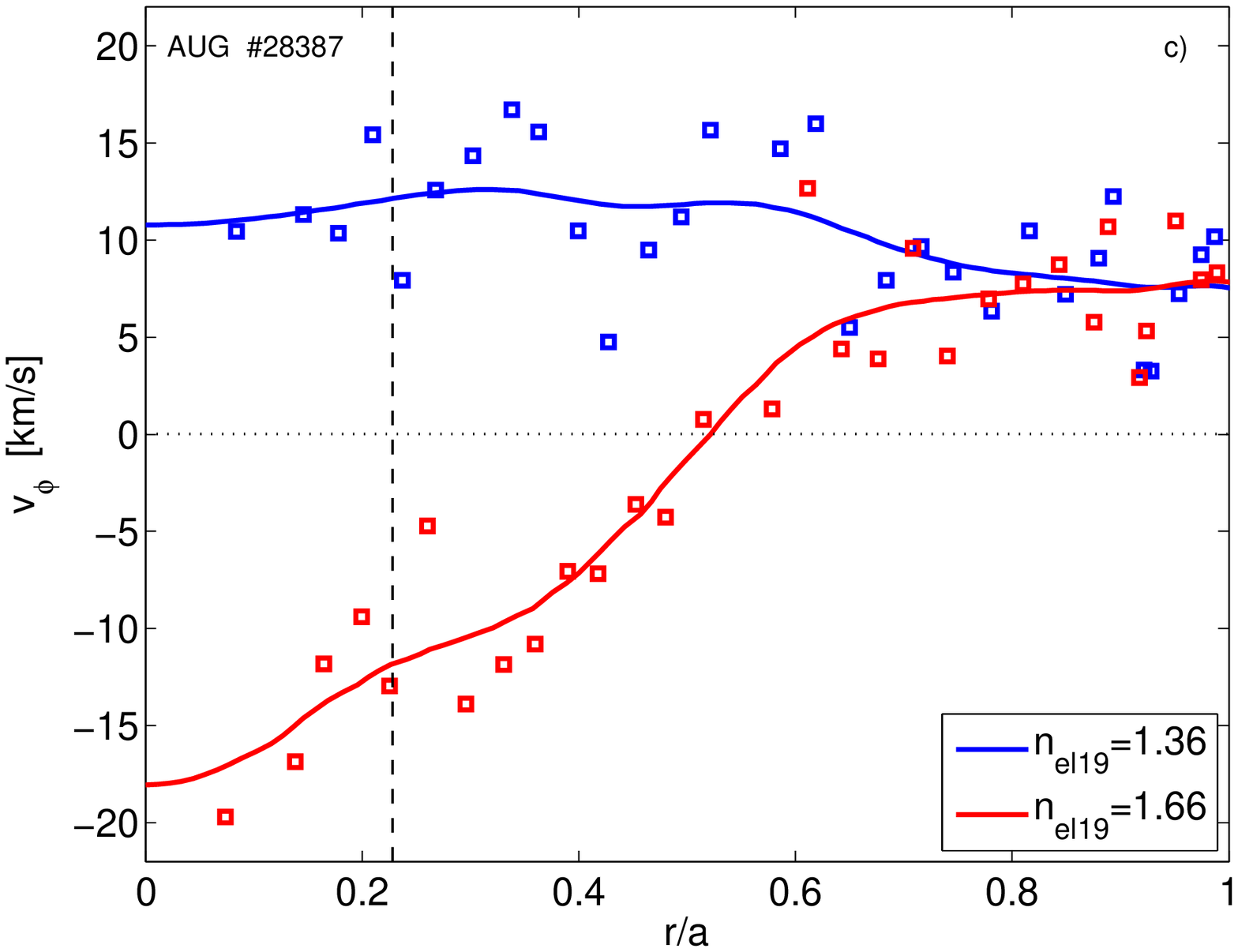}%
        \includegraphics[width=0.32\textwidth]{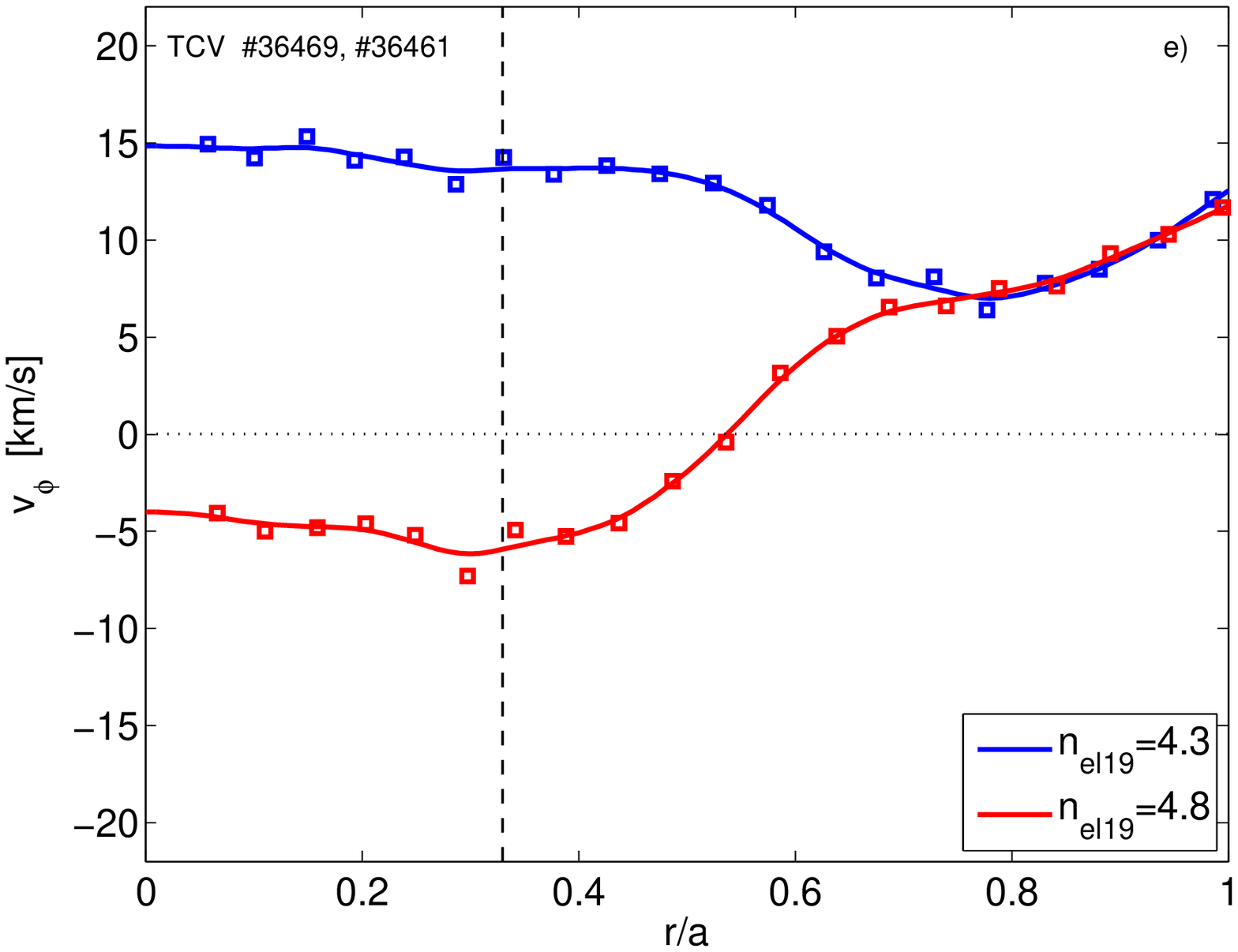}\\%
        \includegraphics[width=0.32\textwidth]{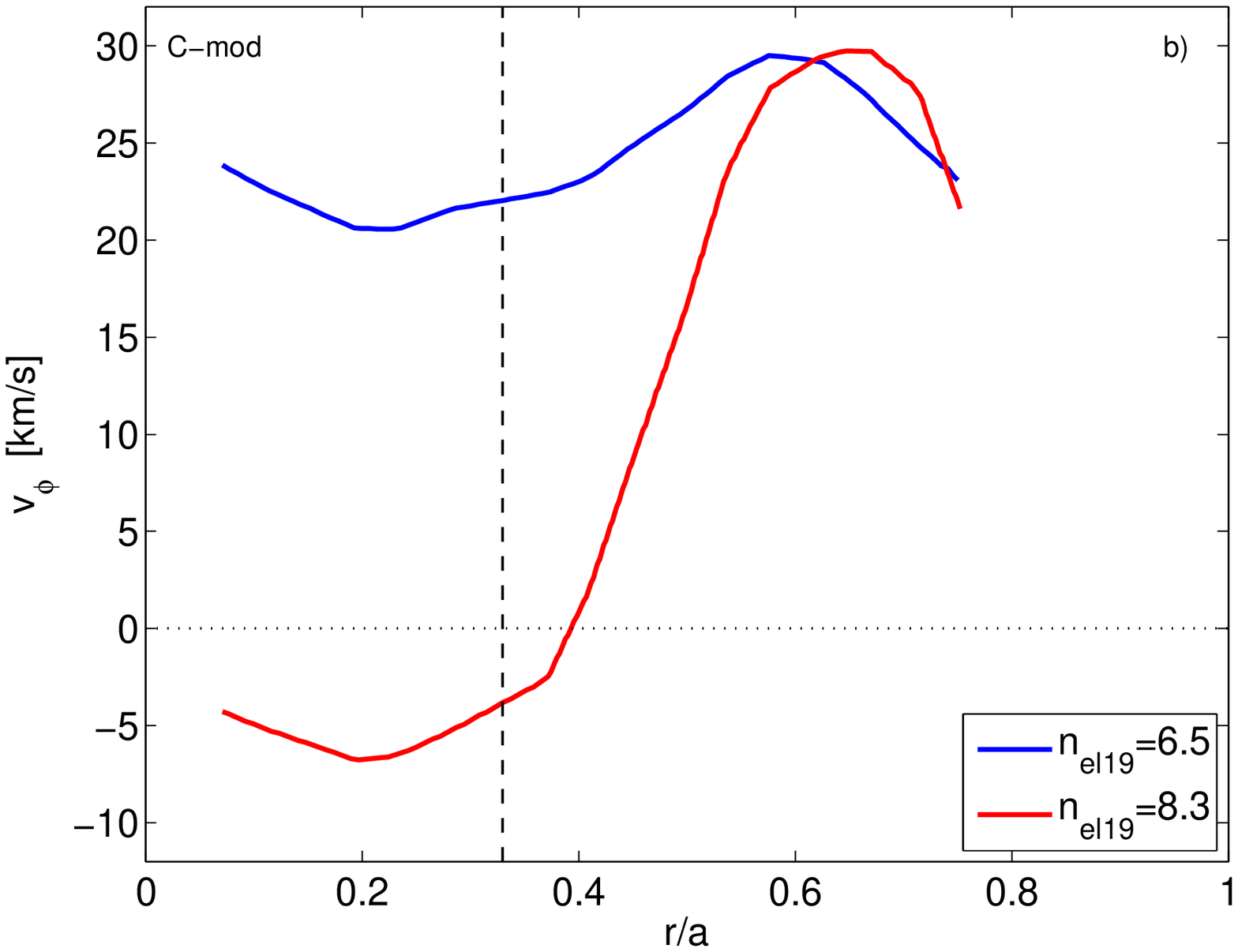}%
        \includegraphics[width=0.32\textwidth]{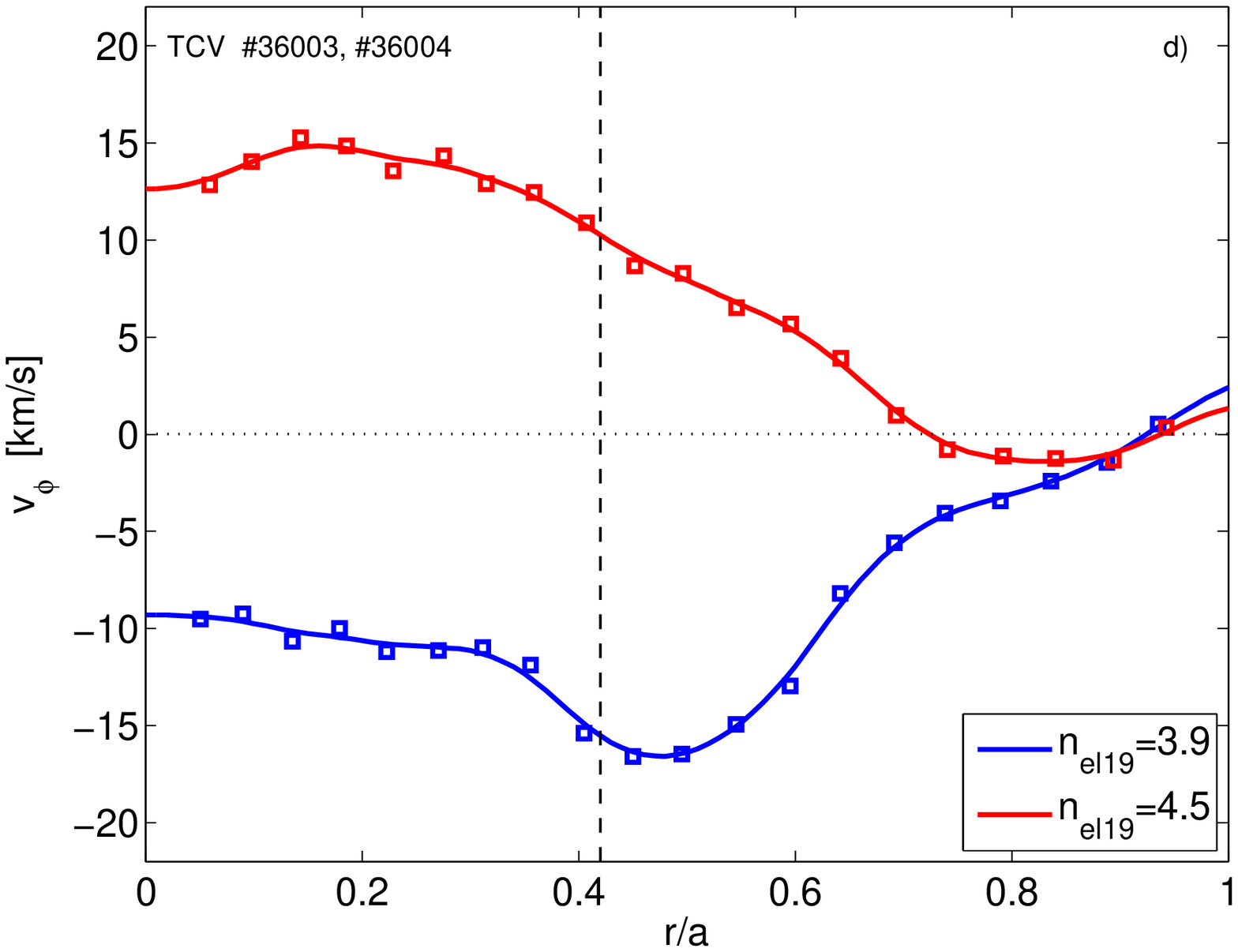}%
        \includegraphics[width=0.32\textwidth]{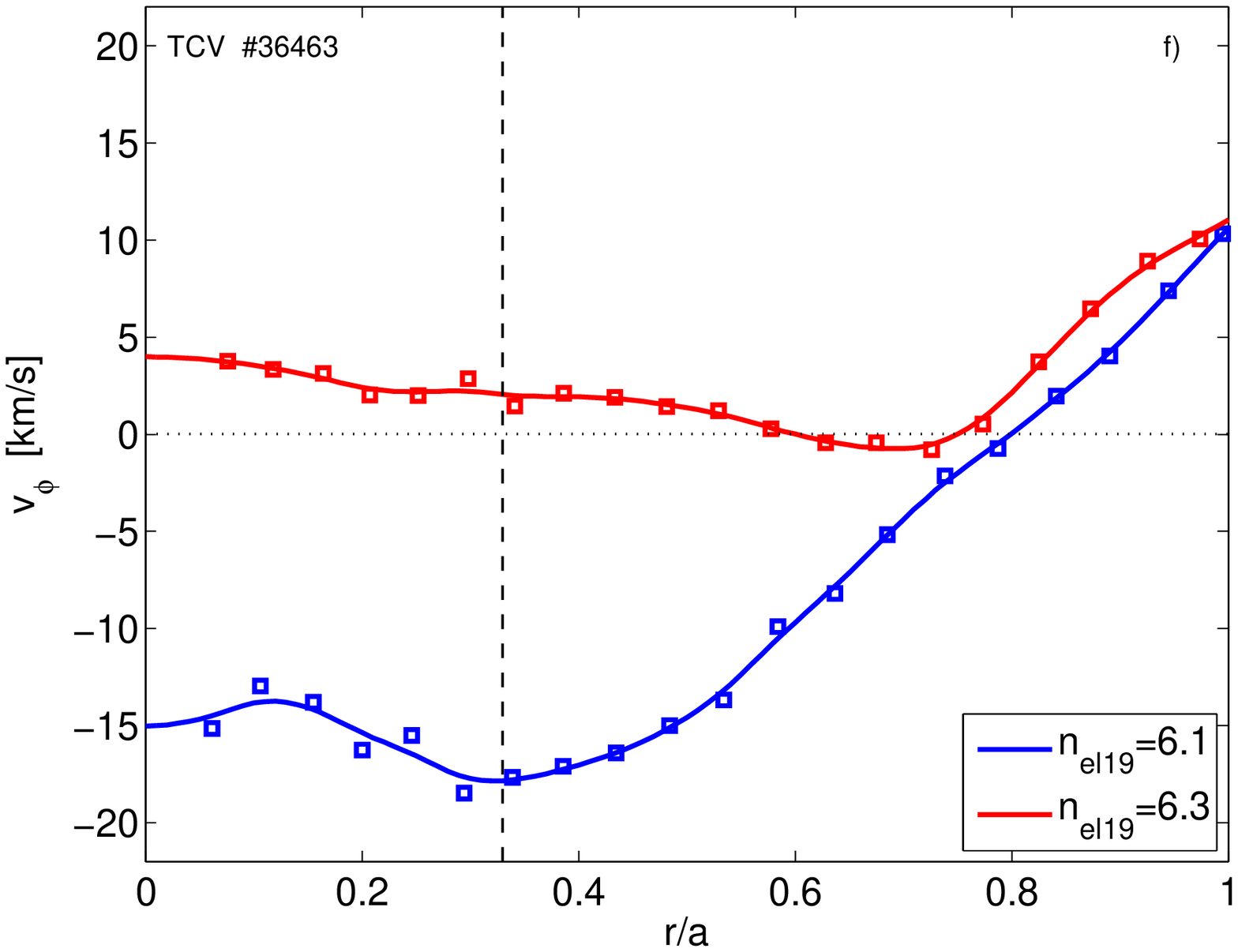}%
        \end{center}
        \caption{\label{fig1} Typical pre/post reversal toroidal rotation profiles measured in C-Mod (adapted from Fig. 13 and 15 in \cite{Rice:NF11}), AUG (adapted from Fig 4 in \cite{McDermott:NF14}) and TCV (adapted from Fig. 6.3 and 8.1 in \cite{Bortolon:PhD09}). The line averaged density for each case is given in units of $10^{19}\,{\rm m}^{-3}$ and the inversion radius is indicated by a dashed line.\\
a) C-Mod, Type I reversal, lower single null ($B\times\nabla B $ upward), $I_p=1.05\,{\rm MA}$, $B_T=5.4\,{\rm T}$, $q_{95}=3.2$.\\
b) C-Mod, Type I reversal, upper single null ($B\times\nabla B $ upward), $I_p=0.8\,{\rm MA}$, $B_T=5.4\,{\rm T}$, $q_{95}=4.7$.\\
c) AUG, Type I reversal, lower single null ($B\times\nabla B $ downward), $I_p=0.5\,{\rm MA}$, $B_T=1.5\,{\rm T}$, $q_{95}=4.9$.\\
d) TCV, Type II.a reversal, limited configuration, $I_p=0.34\,{\rm MA}$, $B_T=1.45\,{\rm T}$, $q_{95}=2.7$.\\
e) TCV, Type I reversal, lower single null ($B\times\nabla B $ downward), $I_p=0.26\,{\rm MA}$, $B_T=1.45\,{\rm T}$, $q_{95}=3.6$.\\
f) TCV, Type II.b reversal, lower single null ($B\times\nabla B $ downward), $I_p=0.26\,{\rm MA}$, $B_T=1.45\,{\rm T}$, $q_{95}=3.6$.}
\end{figure*}

Typical pre- and post-reversal toroidal rotation profiles are shown in Fig.~\ref{fig1} for AUG, C-Mod and TCV plasmas.
\begin{enumerate}
\item In the sawtooth region, the toroidal rotation profile is mostly flat with a bulge in the co-current direction, independent of the reversal state (measurement integrated over several sawtooth cycles)
\item In the gradient region, the toroidal rotation gradient has a wide range of values and often a different sign before and after the reversal. 
\item In the edge region, the toroidal rotation profiles are similar just before and just after the reversal. 
\item In C-Mod, the modification of the rotation profile in Type I reversals occurs in the region $q \lesssim 3/2$.\\
{\footnotesize C-Mod: Fig 16 in \cite{Rice:NF13}}
\end{enumerate}

\subsection{Dynamics} \label{sec:dynamics}
The dynamics of reversals have been investigated during density ramp experiments for Type I reversals in C-Mod and Type II.a reversals in TCV. A similar behaviour is reported in the two cases.
\begin{enumerate}
\item In C-Mod and TCV, the reversal process appears as a clear break in slope of the toroidal rotation response to an increase in density.\\
{\footnotesize C-Mod: Fig. 1 in \cite{Rice:NF11}, TCV: Fig. 5.3 in \cite{Bortolon:PhD09} }
\item \label{exp:bif4} After the reversal process commences, the temporal dynamics of the central toroidal rotation is rather well described by an exponential fit of the form $\exp{[-t/\tau_{\rm rev}]}$. The characteristic time of the reversal $\tau_{rev}$ is comparable to, or longer than, the energy confinement time. In TCV, for the Type II.a reversals shown in Fig. 5.3, 5.4 and 5.6 of \cite{Bortolon:PhD09}, $\tau_{\rm rev}$ ranges from $40$ to $120\U{ms}$. 
\item \label{exp:bif2} When the reversal is triggered by a density ramp, the time scale of the reversal is independent of the density ramp-rate.\\
{\footnotesize C-Mod: Fig. 2 in \cite{Rice:NF11}, TCV: Fig. 5.4 in \cite{Bortolon:PhD09} }
\item \label{exp:bif3} Even with small increases of the line averaged density, stabilising the profiles between the initial and final states of a reversal has never been demonstrated. Fig. 6 in \cite{Duval:PPCF07} shows the typical gap between the two stationary states.
\item During the reversal, there is a transient evolution of the edge rotation ($0.8\lesssim r/a \lesssim 1$) in the direction opposite to that of the core (edge recoil). The edge rotation then relaxes to its pre-reversal value.\\
{\footnotesize C-Mod: Fig. 19, 20 in \cite{Rice:NF11}, TCV: Fig. 5.7 and 5.8 in \cite{Bortolon:PhD09}}
\item \label{exp:bif5} The reversal is, itself, reversible, with some hysteresis in density $>10\%$. There may also be a hysteresis in the plasma current, but this is not as clear due to a relatively slow current diffusion time.\\
{\footnotesize C-Mod: Fig. 4 and 7 in \cite{Rice:NF11}, TCV: Fig. 5.5 in \cite{Bortolon:PhD09}}
\end{enumerate}

\subsection{Critical density for the reversal}
\begin{enumerate}
\item The density threshold for Type I reversals increases with increasing plasma current (C-Mod). This was also indirectly observed in experiments where the plasma current was scanned at constant density (TCV).\\
{\footnotesize C-Mod: Fig. 10 in \cite{Rice:NF11}, TCV: Fig. 6.5 in \cite{Bortolon:PhD09} and Fig. 1 in \cite{Sauter:IAEA10}.}
\item  The density threshold for Type I reversals decreases with increasing toroidal magnetic field.\\
{\footnotesize C-Mod: Fig. 12 in \cite{Rice:NF11}.}
\item In C-Mod, the $I_p$ and $B_T$ dependencies of the density threshold can be unified by a critical density proportional to $1/q_{95}$. For the cases investigated, a critical collisionality of the form $\nu_{\rm rev}\propto n_eZ_{\rm eff}/T_e^2$ works equally well as the factor $Z_{\rm eff}/T_e^2$ was nearly constant at the reversal \cite{Rice:NF13}. In AUG, a critical collisionality better unifies the data than a critical density, see Fig 7 in \cite{McDermott:NF14}.
\item The density threshold for Type II.a reversals (backwards with respect to Type I reversals) decreases with increasing plasma current and, therefore, has an opposite dependence on $I_p$ compared to Type I reversals.\\
{\footnotesize TCV: Fig. 5.6 in \cite{Bortolon:PhD09}.}
\item The density threshold for Type II.a reversals increases with increasing ECH power.\\
{\footnotesize TCV: Fig. 5.17 and 5.18 in \cite{Bortolon:PhD09}.}
\item The scaling of the density threshold for Type II.b and Type II.c reversals with respect to plasmas parameters is, to date, unexplored.
\end{enumerate}

\subsection{Poloidal rotation}
Poloidal rotation profiles before and after Type II.a reversals have been measured in TCV.
\begin{enumerate}
\item The pre- and post- reversal profiles are similar and no large excursions of poloidal rotation are observed during the reversal (on the measurement timescale).\\
{\footnotesize TCV: Fig 5.9 in \cite{Bortolon:PhD09} and Fig 8 in \cite{Duval:PPCF07}.}
\item No departure from the neoclassical theory prediction is observed within the measurement uncertainties ($\lesssim 2\U{km/s}$) \cite{Bortolon:NF13}. 
\end{enumerate}

\subsection{Plasma shape}
\begin{enumerate}
\item In TCV, Type II.a reversals vanished for negative triangularity: the rotation profile is peaked even at low density (with the edge counter-current rotating) and shifts rigidly towards more counter-current rotation when the density is increased \cite{Duval:PoP08}.
\item In KSTAR, no sharp evolution of the toroidal rotation is observed  when ramping the plasma density in low elongation limited plasmas. The toroidal rotation remains counter-current and displays a mild U-curve behaviour with density \cite{Na:NF16}. 
\end{enumerate}

\subsection{Sawteeth and MHD activity} \label{sec:mhd}
\begin{enumerate}
\item Toroidal rotation reversals are often observed in plasmas that also display sawtooth phenomena. However, the presence of sawteeth appears to be a consequence of the constrained operational space. Reversals have been triggered in plasmas exhibiting a wide variety of sawtooth characteristics and no correlation has been established between the reversals and the sawtooth frequency, amplitude or inversion radius \cite{Duval:PoP08}. In AUG, hollow rotation profiles with the core rotating in the counter-current direction (high density branch of Type I reversals) were observed in the absence of detectable sawteeth \cite{McDermott:blabla2016}. 
\item Low amplitude MHD activity is sometimes observed during toroidal rotation reversal experiments. For instance, in TCV, a $(2,1)$ mode and a $(1,1)$ sawtooth precursor, whose amplitude increases with the plasma density is often detected by magnetic probes. As for the sawteeth, no clear correlation is found between the MHD activity and the rotation direction, with co- and counter- rotation observed for similar MHD spectrograms \cite{Duval:PoP08,Bortolon:PhD09}
\end{enumerate}

\subsection{LOC/SOC transition and turbulence changes}
\begin{enumerate}
\item Type I reversals generally occur close to, but not necessarily at, the transition from linear Ohmic confinement (LOC) to saturated Ohmic confinement (SOC) \cite{McDermott:NF14, Na:NF16, Angioni:PRL11, Rice:NF11,Rice:PRL11b, Rice:PoP12, Rice:NF13, Lebschy:EPS16} and to the non-local heat/cold pulse cut-off \cite{Rice:NF13}.
\item The prediction of the turbulence regime from linear stability calculation is a delicate exercise due to the sensitivity of the TEM/ITG transition to input parameters that are difficult to measure precisely (temperature, density and rotation gradients, collisionality, magnetic shear, etc.) and to the choice of the collision operator in the numerical simulations \cite{Manas:PoP15}. In addition, the linear stability does not necessarily reflect the non-linear state. Experimentally, the characterisation of the turbulence regime from temperature and density fluctuation measurements is not straigtforward either. These caveats in mind, linear stability calculations \cite{Angioni:PRL11, White:PoP13, McDermott:NF14} for AUG and C-Mod plasmas indicate that toroidal rotation reversals often occur close to the boundary between the TEM and ITG instabilities and fluctuations measurements on C-Mod \cite{Rice:NF11,Rice:NF13,White:PoP13} show changes in the fluctuation spectra across toroidal rotation reversals. These changes in the turbulence characteristics do not appear, however, to trigger the reversal as they can occur in a region where the toroidal rotation gradient is not modified \cite{White:PoP13} or the rotation profile experience a reversal without a change of the predicted dominant instability \cite{McDermott:NF14}. 
\end{enumerate}

\subsection{Dependence of the toroidal rotation gradient on plasmas parameters}
\begin{enumerate}
\item Multi-variable regressions performed for a large database of AUG Ohmic L-modes \cite{McDermott:NF14} and for a reduced set of  AUG and TCV Ohmic L-modes \cite{McDermott:EPS15}, show that the toroidal rotation gradient is mostly correlated to the normalised density gradient $R/L_n$ and effective collisionality $\nu_{\rm eff}$. Larger $R/L_n$ increases the hollowness of the rotation profile. Large variations of the toroidal rotation gradient (including a change of sign) can, nevertheless, be observed at nearly constant $R/L_n$ \cite{Lebschy:EPS16}.
\item Interestingly, in AUG the strong dependence of the toroidal rotation gradient on $R/L_n$ is also observed for a wide operational range including Ohmic and electron cyclotron heated L-modes and H-modes \cite{Angioni:PRL11}. Strongly hollow rotation profiles are only observed at large $R/L_n$ values.
\end{enumerate}

\section{Why is momentum transport considered as the key to explain the reversals?} \label{sec:inteprexp}
The mechanisms invoked to explain toroidal rotation reversals need to be consistent with all the experimental observations summarised in Sec.~\ref{sec:expobs}. This includes the observed dynamics (time scale, edge recoil, hysteresis), the parametric dependencies of the critical density and the constancy of the pre/post-reversal rotation profiles in the edge region.

The main mechanisms expected to impact the intrinsic rotation profile in the gradient region, where the reversal takes place, are the neoclassical and turbulent momentum transport (momentum redistribution), the neoclassical toroidal viscosity due to field ripple or a strong MHD mode (damping towards a diamagnetic level offset), a torque due to resonant non-axisymmetric fields (locking to the wall) and sawteeth (momentum redistribution and possibly transient torque).
Sawteeth and strong MHD modes certainly affect the intrinsic rotation profile, but a causal link between sawteeth/MHD and toroidal rotation reversal has yet to be established. As described in Sec.~\ref{sec:mhd}, toroidal rotation reversals are observed with little to no MHD activity and for a variety of  MHD spectrograms and sawtooth behaviour, independent of the reversal state,  
The magnitude of the toroidal rotation in the pre- and post- reversal states is often larger than a diamagnetic level rotation and, therefore, somewhat incompatible with NTV or resonant non-axysimmetric fields as the dominant process. In addition, rotation reversals are observed in tokamaks with very low ripple like KSTAR but not in tokamaks with high ripple like Tore Supra \cite{Bernardo:PPCF15}. In Tore Supra, a slightly hollow profile of counter-current rotation is measured in Ohmic L-modes that is satisfactorily described by NTV theory. Close to the LOC/SOC transition, a small departure from NTV predictions is observed, reminiscent of a toroidal rotation reversal but far from significantly affecting the rotation profile: when NTV dominates, toroidal rotation reversals are hampered \cite{Bernardo:PPCF15}. %
Summing up these various considerations, momentum transport is the only viable candidate left to explain rotation reversals. An additional strong argument in this direction is brought by the transient acceleration of the plasma edge in the direction opposite to that of the core observed during a reversal (edge recoil, see Sec.~\ref{sec:dynamics}). The edge recoil cannot easily be produced by a localised torque or damping term: this would require us to invoke several radially localised contributions of opposite directions and different temporal behaviour. In contrast, a sudden (i.e. faster that the momentum confinement time) change of momentum transport in the plasma core at constant edge momentum transport produces an edge recoil as a consequence of momentum conservation. The profile evolves on a time scale dictated by momentum diffusion, i.e. $\tau_{\rm rev}\sim a^2/\chi_\varphi$ with $\chi_\varphi$ the momentum diffusivity.

\section{Momentum transport theory} \label{sec:theory}
\subsection{Momentum conservation and momentum flux}
Assuming  momentum transport to be the only mechanism at play, the toroidal rotation in the core of an axisymmetric tokamak is governed by the redistribution of toroidal angular momentum, e.g. \cite{Callen:PoP09}:
\begin{equation}\label{eq:1Dmom}
\frac{\partial}{\partial t}   \sum_s \left< n_s m_s R v_{\varphi,s} \right> + \frac{1}{V'}\frac{\partial}{\partial r} \left[ V' \Pi_\varphi \right] = 0
\end{equation}
with $n_s$, $m_s$ and $v_{\varphi,s}$ the density, mass and toroidal velocity of species $s$, respectively, $R$ the local major radius, $\left<.\right>$ the flux surface average, $r$ a radial coordinate (flux surface label), $V$ the flux surface volume, $V'=\partial V /\partial r$ (radial derivative) and $\Pi_\varphi=\left<\mathbf{\Pi}_\varphi\cdot \nabla r\right>$ the flux surface averaged radial component of the toroidal momentum density flux. 
Eq.~(\ref{eq:1Dmom}) is obtained from the flux surface average of the momentum conservation equation and simply states that, in the absence of sources and sinks, the evolution of the toroidal angular momentum density is driven by the divergence of the momentum flux $\Pi_\varphi$. 

Up to first order in $\rho_*=\rho_i/R_0$, the species flow entering Eq.~(\ref{eq:1Dmom}) lies within a flux surface and is given by the sum of the parallel streaming along the magnetic field lines, of the $E\times B$ drift and of the diamagnetic flow 
\begin{equation} \label{eq:flow}
\mathbf{v}_s = v_\parallel \mathbf{b} + \mathbf{v}_{\rm E} + \mathbf{v}_{\rm dia} + \mathcal{O}(\rho_*^2)
\end{equation}
Here, $\rho_i=m_iv_{\rm thi}/(eB_0)$ is the main ion Larmor radius, $R_0$ and $B_0$ are a reference major radius and magnetic field, respectively,  and $v_{\rm thi}=\sqrt{2T_i/m_i}$ is the thermal velocity. 
The parallel flow can be split into three components $v_\parallel=v_\parallel^{\rm E}+v_\parallel^{\rm dia}+v_\parallel^\theta$ so that $v_\parallel^{\rm E}\mathbf{b} +  \mathbf{v}_{\rm E}$ and $v_\parallel^{\rm dia}\mathbf{b} +  \mathbf{v}_{\rm dia}$ are purely toroidal whereas the remaining contribution to the total flow, $v_\parallel^\theta \mathbf{b}$, has finite poloidal and toroidal components. The two purely toroidal flows are given by:
\begin{eqnarray} \label{eq:exbflow}
 v_\parallel^{\rm E}\mathbf{b} +  \mathbf{v}_{\rm E} &=& R\omega_\Phi  \mathbf{e}_\varphi =- R \frac{\partial \Phi}{\partial \psi} \mathbf{e}_\varphi \\
 v_\parallel^{\rm dia}\mathbf{b} +  \mathbf{v}_{\rm dia} &=& R \omega_{p,s} \mathbf{e}_\varphi = - R \frac{1}{Z_s en_s}\frac{\partial p_s}{\partial \psi} \mathbf{e}_\varphi \label{eq:diaflow}
\end{eqnarray}
with $\Phi$ the electrostatic potential, $\psi$ the poloidal magnetic flux, $\mathbf{e}_\varphi$ the unit vector in the toroidal direction and $Z_s$ and $p_s$ the species charge number and pressure, respectively. The parallel flow $v_\parallel^\theta$ is constrained by neoclassical physics.
Combining Eqs.~(\ref{eq:flow}), (\ref{eq:exbflow}) and (\ref{eq:diaflow}), yields the customary expression of the toroidal flow \cite{Hirshman:NF81}:
\begin{equation} \label{eq:flowstd}
v_{\varphi,s} = R\omega_\Phi + R\omega_{p,s} + v_{\theta,s}\frac{B_t}{B_p}
\end{equation}
The first term, related to the $E\times B$ flow, is the lowest order contribution. It is species independent and can assume arbitrarily large values in an axisymmetric tokamak. The lowest order momentum transport theory is formulated with respect to $\omega_\Phi$. The next contributions, related to the diamagnetic and poloidal flows, are first order in $\rho_*$ for a neoclassical level poloidal flow and roughly scale as $\frac{1}{2}\rho_*\frac{B_t}{B_p}v_{\rm thi}R/L_{T_i}$, with $R/L_{T_i}=-R_0\partial \ln T_i/\partial r$ the normalised temperature gradient. For $\rho_*=1/600$, $B_t/B_p=10$ and $R/L_{T_i}=6$, the first order toroidal flow in Eq.~(\ref{eq:flowstd}) is about $0.05 \,v_{\rm thi}$ and, therefore, not negligible compared to the total toroidal flow, which is often less than $0.2 \,v_{\rm thi}$ for intrinsic rotation. When dealing with intrinsic rotation, the distinction between the total toroidal flow $v_{\varphi,s}$, that is the measured quantity, and the lowest order toroidal flow $R\omega_\Phi$ therefore needs to be taken into account. 

The momentum flux entering the transport equation, Eq.~(\ref{eq:1Dmom}), is now decomposed into diagonal (diffusive), pinch (convective) and residual stress components:
\begin{equation}
 \Pi_\varphi = n m R_0 v_{\rm thi}\left[ \chi_{\varphi} u' + R_0V_{\varphi}u + C_{\varphi} \right]
\end{equation}
Here, the decomposition is performed with respect to the lowest order flow, i.e. the diagonal part components with respect to $u'=- R_0/v_{\rm thi}\partial \omega_\Phi/\partial r$ and the pinch components with respect to $u=R_0\omega_\Phi/v_{\rm thi}$. 
In the expression above, $nm=\sum n_sm_s$ is the species averaged mass density and the momentum transport coefficients have also been species averaged using $A = \sum n_s m_s A_s/\sum n_s m_s$ where $A$ represents the momentum diffusivity $\chi_\phi$, pinch velocity $V_\varphi$ or residual stress coefficient $C_{\varphi}$. 
In stationary state, $\Pi_\varphi=0$ so the intrinsic rotation profile is determined by the balance between the diagonal flux, which tends to flatten the profile, and the non-diagonal flux (pinch and residual stress) which tends to sustain a finite gradient. The sign and magnitude of the resulting rotation gradient is dictated by the ratio of the pinch and residual stress components to the momentum diffusivity: 
\begin{equation}
 u' = - \frac{R_0V_\varphi}{\chi_\varphi}u - \frac{C_\varphi}{\chi_\varphi}
\end{equation}
The fundamental difference between the pinch and residual stress is that only the pinch requires a finite rotation to sustain a gradient. A residual stress contribution is, therefore, required to describe rotation profiles crossing zero, as observed in Fig.~\ref{fig1} for the impurity rotation $v_{\varphi,s}$ or in \cite{McDermott:NF14} for the $E\times B$ angular frequency $\omega_\Phi$.

In the core of an axisymmetric tokamak, the neoclassical momentum flux is typically an order of magnitude smaller than the gyro-Bohm momentum flux \cite{Belli:PPCF09} and negligible compared to the turbulent momentum flux. The following discussion, therefore, focuses on turbulent momentum transport. The main mechanisms are briefly outlined in the framework of gyrokinetic theory, emphasising their potential link with rotation reversals. For a more comprehensive description of the theory of turbulent momentum transport and further references to the original work, the reader is referred to published reviews \cite{Peeters:NF11,Angioni:NF12,Diamond:NF13,Parra:PPCF15}.

\subsection{Lowest order contributions}
To lowest order, with respect to the gyrokinetic ordering (local limit, $\rho_*\rightarrow 0$), five mechanisms that can generate a momentum flux are described. The parallel \cite{Mattor:PoF88} and perpendicular \cite{Dominguez:PoFB13} components of the toroidal flow shear give rise to a diagonal flux. For positive magnetic shear, these two contributions have opposite sign and the perpendicular component of the toroidal flow shear acts to reduce toroidal momentum diffusivity \cite{Casson:PoP09}. The pinch also has two contributions: the Coriolis pinch \cite{Peeters:PRL07} and the momentum carried by any particle flux. In the stationary state, the second contribution vanishes if no particle source remains.
Finally, the only contribution to the lowest order residual stress arises from the up-down asymmetry of the magnetic flux surfaces $C_\varphi^{\rm FS}$  \cite{Camenen:PRL09}.

The ratio of the toroidal momentum diffusivity and ion heat diffusivity, the Prandtl number, is typically predicted as $\textrm{Pr} = \chi_\varphi/\chi_i\sim 0.7$ \cite{Peeters:NF11}, but values in the range $0.4$ to $1.5$ are possible depending on the plasma parameters \cite{Casson:PoP09}. The Coriolis pinch is generally directed inward and acts to increase the absolute value of the rotation. The pinch to diffusivity ratio $R_0V_\varphi/\chi_\varphi$ typically ranges from -1 to -4 with a marked dependence on the normalised density gradient $R/L_n$. The Coriolis pinch tends to be smaller in the TEM regime \cite{Kluy:PoP09} and can be directed outward close to the kinetic ballooning mode threshold \cite{Hein:PoP11}. The ratio of the residual stress from the flux surface asymmetry to the momentum diffusivity is typically $|C_\varphi^{\rm FS}/\chi_\varphi|\lesssim 1$ near the edge where the flux surface shaping is the highest and  $|C_\varphi^{\rm FS}/\chi_\varphi|\lesssim 0.3$ in the core \cite{Camenen:PRL09, Ball:PPCF16}. The sign of $C_\varphi^{\rm FS}$ is determined by the flux surface asymmetry and the direction of the magnetic field. 
The momentum diffusivity, pinch and up-down asymmetry residual stress have all been identified experimentally and found to be in fair agreement with lowest order gyrokinetic theory predictions \cite{Solomon:PRL08, Weisen:NF11, Tala:NF11, Camenen:PRL10}.
As the intrinsic rotation gradient in the vicinity of $u=0$ is typically between $-1.5\lesssim u' \lesssim 1.5$, including up-down symmetric plasmas for which $C_\varphi^{\rm FS}=0$, intrinsic rotation can clearly not be described by the lowest order theory: it lacks residual stress contributions. 

\subsection{First order contributions}
To next order in $\rho_*$, new contributions to the residual stress arise from:
\begin{enumerate}
\item the impact of a poloidally inhomogeneous turbulence on the parallel symmetry \cite{Sung:PoP13}
\item profile shearing, i.e. the shear in the drifts and parallel motion due to first and second order derivatives of the magnetic equilibrium, density and temperature profiles \cite{Waltz:PoP11,Camenen:NF11,Buchholz:PoP14}
\item the generic impact of a radially inhomogeneous turbulence on the parallel symmetry \cite{Diamond:PoP08,Gurcan:PoP10} 
\item the impact of a radially inhomogeneous turbulence on passing ions with different orbit shifts \cite{Stoltzfus:PRL12}
\item the deviation of the equilibrium distribution function from a Maxwellian, i.e. the impact of the neoclassical equilibrium \cite{Barnes:PRL13}, which includes the pressure gradient contributions to the $E\times B$ shear \cite{Gurcan:PoP07}. 
\end{enumerate}
For the parameter dependencies, all contributions that rely on coupling by parallel compression between density and parallel velocity fluctuations increase in magnitude with $R/L_n$. This is the case for contributions (i-iii) and a part of (v). The dependence already appears in reduced fluid models when considering a generic parallel symmetry breaking, see e.g. \cite{Peeters:PoP09a}. Another robust feature of residual stress is that its magnitude tends to be smaller for TEM dominated turbulence than for ITG dominated turbulence, typically by a factor $\gtrsim 2$, consistently with the more symmetric mode structure with respect to the horizontal midplane obtained in the TEM regime.
The residual stress contributions related to the radial inhomogeneity of turbulence \cite{Gurcan:PoP10,Stoltzfus:PRL12}, to the shear in the perturbed $E\times B$ drift advection of the background \cite{Waltz:PoP11} and to the neoclassical equilibrium \cite{Barnes:PRL13} all strongly depend on the second order derivatives of the temperature and/or density profiles. This dependence can extend as far as to change their sign. The neoclassical equilibrium residual stress also strongly depends on, and can change sign with, the ion-ion collisionality. The contribution related to the shear of the parallel motion and of the curvature and $\nabla B$ drift \cite{Camenen:NF11} does not depend on second order derivatives. 
Interestingly, contribution (iv) alone strongly depends on the radial position of the X-point. Its impact on the rotation gradient is limited to the edge region \cite{Stoltzfus:PRL15} and therefore not directly relevant for toroidal rotation reversals.\\
Of course, for all first order contributions, $C_\varphi/\chi_\varphi$ is expected to depend on $\rho_*$. 
This dependence is linear in $\rho_*$ for (iii) and (v) according to analytical calculations, but may be more complicated for the other contributions. For instance, it has been shown in global non-linear simulations that for profile shearing, $C_\varphi/\chi_\varphi$ is first linear in $\rho_*$ but saturates at high $\rho_*$ values \cite{Buchholz:PoP14}. Overall, the $\rho_*$ scaling of residual stress is still debated and deserves further investigation. What is certainly true, however, is that the exponent on any $\rho_*$ scaling is between 0 and 1 and could quite possibly depend on the plasma parameters. 

\subsection{Numerical simulations}
All the contributions listed above combine to generate the total residual stress. Unfortunately, many tend to be of comparable magnitude, at least from simple scaling arguments, with various signs, making the prediction of their sum a delicate exercise. A quantitative prediction of intrinsic rotation therefore requires numerical simulations.
The lowest order momentum flux can be computed in gyrokinetic $\delta f$ flux-tube codes provided that the background $E\times B$ toroidal flow and an arbitrary flux surface geometry are included. 
The first order contribution (i) can also be computed within the flux-tube approach but requires the inclusion of higher order parallel derivatives. 
The contributions (ii) to (iv) require a radially global approach.
Contribution (v) can be treated in a $\delta f$ flux-tube code by adding the neoclassical correction to the background Maxwellian distribution function. It can also be treated by solving the coupled neoclassical and turbulent problem. The second option requires an accurate collision operator and is considerably more computationally expensive as the simulations must cover several ion-ion collision times to reach a stationary state with respect to the neoclassical physics. This method includes, however, the impact of turbulence on neoclassical physics, which may also be relevant for momentum transport \cite{Idomura:PoP14}.
Finally, all simulations that aim for a quantitative prediction of the momentum flux must treat the electrons kinetically as an adiabatic electron approximation has a dramatic impact on the parallel symmetry \cite{Peeters:PoP09b}. 

The first principle prediction of the intrinsic rotation profile resulting from momentum transport including the interplay between neoclassical and turbulence physics represents a formidable challenge. Only one example of a global simulation with kinetic electrons (at reduced ion to electron mass ratio) including all the lowest and first order contributions to momentum transport and evolving the rotation profile over a confinement time has been reported \cite{Idomura:JCP16}. It remains an extremely important result as it demonstrates that a rotation profile reaching $0.15 v_{\rm thi}$ can be sustained by the internal redistribution of momentum by turbulence, see Fig~10 in \cite{Idomura:JCP16}, lending further support to the interpretation of intrinsic rotation in this framework.
It also confirms the critical role of including kinetic electrons as the toroidal rotation was shown to develop in the opposite direction when the electrons were described adiabatically.

At some point, the issue was raised as to whether the conventional gyrokinetic ordering was sufficient to properly describe turbulent momentum transport, in particular in full-f global simulations \cite{Parra:PoP10}. It was first proven in the context of the gyrokinetic field theory that momentum conservation is guaranteed to any order provided the approximations are made at the level of the Hamiltonian \cite{Scott:PoP10} and it was then demonstrated numerically that the conventional gyrokinetic ordering is sufficient to describe momentum transport in the long wavelength approximation that is valid for ITG and TEM turbulence \cite{Idomura:CSD12}.  

\section{On-going modelling activities} \label{sec:modelling}
At present, numerical simulations as reported in \cite{Idomura:JCP16} are far too costly to be systematically compared to experimental observations. Modelling activities therefore focus on the residual stress contributions separately in the hope that one of these contributions dominates the others in magnitude. 
The collisionality dependence of the neoclassical equilibrium residual stress $C_\varphi^{\rm NC}$ is appealing and was invoked to explain Type I rotation reversals in MAST \cite{Hillesheim:NF15}. A simplified qualitative model was used to predict the reversal state. According to this model, the rotation profile is predicted to be peaked in the plasma region where the collisionality is lower than a threshold value and hollow in the region above, with the transition between the two regions moving radially inward across a density ramp. While the prediction of the reversal state was reasonably successful, the model did not appear compatible with the experimental observation that the rotation profile is strongly modified at the critical density but relatively independent of the density before and after the reversal. 
More recently, the impact of $C_\varphi^{\rm NC}$ has been investigated for the AUG database assembled in \cite{McDermott:NF14} and covering pre- and post- reversal profiles. The focus of this work was the prediction of the toroidal rotation gradient around mid-radius with the modelling based on a quasi-linear approach supported by a few non-linear simulations \cite{Hornsby:NF16}. The gyrokinetic simulations included all the lowest order terms and the residual stress driven by the neoclassical equilibrium. The latter appeared comparable in magnitude to the up-down asymmetry residual stress and the Coriolis pinch. The predicted toroidal rotation gradient was up to an order of magnitude smaller than measured, demonstrating the need to invoke other contributions to explain the measurements. 
The impact of profile shearing was investigated for a DIII-D plasma in which the toroidal rotation was nearly zeroed out by counter-current neutral beam injection \cite{Waltz:PoP11}. In these conditions, the Coriolis pinch is negligible and the momentum input balances the residual stress. The simulations were performed in the non-linear regime, including all the lowest order contributions to the momentum flux and the profile shearing residual stress (from first and second order derivatives). Around mid-radius, the predicted momentum flux was comparable in magnitude to experiment, but could differ, even in sign, depending on the chosen second order derivatives of the temperature and density profiles. Further studies are required to better quantify the relevance of this contribution. 
Finally, the impact of a generic symmetry breaking term was explored for the AUG databases of \cite{McDermott:NF14} and \cite{Angioni:PRL11} by imposing a finite ballooning angle shift $\theta_0$ in the linear gyrokinetic simulations. A unique $\theta_0$ value was chosen for all the cases in the TEM regime and another for those in the ITG regime. After these two ad-hoc values are chosen, the experimental toroidal rotation gradient around mid-radius is surprisingly well reproduced by the quasi-linear prediction across the whole database capturing the strong $R/L_n$ dependence of the rotation gradient. This suggests that the residual stress mechanism sustaining the intrinsic rotation profile relies on the coupling, by parallel compression, between density and parallel velocity fluctuations, as this coupling directly engenders a $R/L_n$ dependence of residual stress.

\section{Summary and discussion} \label{sec:discussion}
Based on the experimental observations gathered in the last ten years in AUG, C-Mod, MAST, KSTAR and TCV and the developments in the theory of intrinsic rotation, turbulent momentum transport appears to be the most likely candidate to explain toroidal rotation reversals in the core of Ohmic L-modes.\\
Concerning the stationary rotation profiles, the lowest order contributions in the turbulent momentum flux (the diagonal part, the Coriolis pinch and the up-down asymmetry residual stress) cannot account for the experimental observations and higher order contributions to the residual stress are required. 
The first order contributions are now identified \cite{Parra:PPCF15} and in the process of being tested against the experimental observations. Combining one of these first order contributions, the neoclassical equilibrium residual stress, with the lowest order contributions was recently demonstrated to be insufficient to reproduce the experimentally measured intrinsic rotation gradient for a database of AUG Ohmic L-modes \cite{Hornsby:NF16}.
The focus is now moving to another first order contribution, profile shearing residual stress, which was shown to be sufficiently large to reproduce the required momentum flux in a zeroed-rotation DIII-D case \cite{Waltz:PoP11}. One difficulty of this validation exercise arises from the dependence of several residual stress contributions on the second derivatives of the temperature and density profiles, which are unlikely to be ever sufficiently well measured experimentally. There are two main ways to minimise the impact of this issue: either perform the modelling at multiple radial positions as in \cite{Hornsby:NF16} to account for the constraint engendered by the value of the first derivatives or, when possible, use a sufficiently complete simulation to compare the magnitude of the different residual stress terms.
At present, it remains unclear whether a first order contribution dominates for specific experimental conditions.

For the reversals dynamics, the rotation is observed to be a very sharp function of the plasma density at the reversal, at least in C-Mod and TCV. Such a sharp variation could be the signature of:
\begin{enumerate}
\item a continuous but sharp dependence of $C_\varphi/\chi_\varphi$ on a plasma parameter that varies in the density ramp, e.g. density, collisionality, $T_e/T_i$, etc.
\item a bifurcation in momentum transport triggered by the density increase
\item a moderate dependence of $C_\varphi/\chi_\varphi$ on a plasma parameter that exhibits a strong variation (continuous or bifurcation) close to the critical density
\end{enumerate}
Conceptually, a bifurcation is very different from a continuous transition as it requires unstable states and some direct feedback of the rotation profile on momentum transport. Some features of the reversals in C-mod and TCV (the insensitivity of the dynamics to the density ramp rate, the hysteresis and the gap observed in the stationary profiles) suggest a bifurcation rather than a continuous transition. This aspect would deserve further investigation, in particular in AUG and KSTAR, as the choice between a bifurcation and a continuous transition not only impacts the way data should be handled in multi-variable regressions (one or two sets?) but also provides a strong constraint on the theoretical solution. %
From a theoretical perspective, a mechanism that supports hypothesis (i) is not directly offered by current theories, since the dependence of turbulent transport on plasma parameters is predicted to be rather mild in general. A change of sign of one of the residual stress components at the TEM/ITG transition \cite{Diamond:PoP08,Camenen:NF11} could be invoked but the TEM/ITG transition is, itself, not a particularly sharp function of collisionality (it occurs at a different collisionality for the different wavevectors). For hypothesis (ii), a bifurcation in momentum transport could be triggered if the momentum diffusivity becomes locally negative, i.e. the momentum flux locally decreases as the rotation gradient increases. This could, in principle, occur if the contribution to the toroidal momentum diffusivity from the perpendicular dynamics overcomes the parallel one, which requires large values of $\epsilon/q$.  Whether such a mechanism can be at play for realistic plasma conditions remains, however, to be demonstrated. Hypothesis (iii) can probably be dismissed as no plasma parameter except the toroidal rotation has so far been observed to strongly vary at the reversal and this despite an exhaustive search.\\
To summarise, in spite of considerable progress, there is, to date, no modelling that quantitatively predicts the core intrinsic rotation gradient over a large scale database encompassing pre- and post- reversal profiles, nor the dynamics of a reversal. Possible routes to progress are suggested below.

\begin{enumerate}
\item Is there a single or several reversals? Is there a common threshold on a local parameter that unifies the different types of reversals?\\
Important parameters for turbulent transport are the normalised gradients $R/L_{T_e}$, $R/L_{T_i}$ and $R/L_n$, the safety factor, the magnetic shear, $T_e/T_i$, the collisionality, the local plasma $\beta$ and the magnetic equilibrium (elongation, triangularity, etc.).
\item In the same vein, further characterisation of the scalings of the threshold(s) in terms of local plasma parameters, in particular for Type II reversals would be helpful. Here, an important issue is whether a critical collisionality better unifies the data than a critical density. Is that the case in all devices?  Dedicated experiments with electron cyclotron heating power ramps may help decouple density and collisionality effects. 
\item Does the strong correlation between the intrinsic rotation gradient and $R/L_n$ observed in AUG hold across C-Mod, KSTAR and the full TCV databases? This would hint at a residual stress mechanism that relies on parallel compression and would merits examination over as wide an operational range as possible. 
\item An interesting observation from C-Mod is that the region where the toroidal rotation reverses is typically restricted to $q\lesssim 3/2$. In TCV, Type II.a reversals were only observed for a sufficiently low $q_{95}$ value and in KSTAR no reversals are observed for low elongation high $q_{95}$ plasmas, which may or not be related.
From the theory standpoint, the toroidal projection of the perpendicular flow becomes larger at low $q$ and a marked $q$ dependence could suggest an enhanced role of the perpendicular dynamics ($E\times B$ shearing and radial-perpendicular Reynolds and Maxwell stress) close to the reversal. 
Again, a more systematic characterisation of the radial region where the reversal is observed, including different $q_{95}$ values, would bring new elements to this issue.
\item In TCV, toroidal rotation reversals do not occur for negative triangularity (in the sense that a sharp transition is not observed). This should be better understood, in particular whether this is connected to the stabilisation of TEM turbulence at negative triangularity \cite{Marinoni:PPCF09}. More generally, plasma shaping offers a convenient tool to help decouple the plasma current and the edge safety factor, which could be used in limited and diverted plasmas to broaden the operational domain, as in KSTAR experiments \cite{Na:NF16}.
\item Finally, as mentioned in \cite{Angioni:NF12, Diamond:NF13}, there may be a similarity worth investigating between toroidal rotation reversals in Ohmic L-modes and the impact of electron cyclotron heating in H-modes with and without NBI injection \cite{Angioni:PRL11, deGrassie:PoP07, Sommer:NF12, Yoshida:PRL09, McDermott:PPCF11}. \end{enumerate}

\ack
Fruitful discussions with J. Hillesheim and O. Sauter are warmly acknowledged.

\section*{References}
\label{sec:references}

\bibliographystyle{unsrt}
\bibliography{YC_EPS2016}
\end{document}